\documentclass[aps,prl,twocolumn]{revtex4}
\usepackage{graphicx}

\begin{document}

\title{Gravitational lensing as a probe of cold dark matter subhalos}  
\author{Erik Zackrisson and Teresa Riehm}
\affiliation{Oskar Klein Centre for Cosmoparticle Physics, Department of Astronomy, Stockholm University,\\ 10691 Stockholm, Sweden\\ E-mail addresses: ez@astro.su.se, teresa@astro.su.se}
\date{\today}

\begin{abstract}
In the cold dark matter scenario, dark matter halos are assembled hierarchically from smaller subunits. Some of these subunits are disrupted during the merging process, whereas others survive temporarily in the form of subhalos. A long-standing problem with this picture is that the number of subhalos predicted by simulations exceeds the number of luminous dwarf galaxies seen in the the vicinity of large galaxies like the Milky Way. Many of the subhalos must therefore have remained dark or very faint. If cold dark matter subhalos are as common as predicted, gravitational lensing may in principle offer a promising route to detection. In this review, we describe the many ways through which lensing by subhalos can manifest itself, and summarize the results from current efforts to constrain the properties of cold dark matter subhalos using such effects.
%(No more than 150 words)
\end{abstract}
\maketitle

\section*{1. Introduction: Subhalos and satellite galaxies}
The cold dark matter (CDM) model is based on the hypothesis that a significant fraction (around 23\% \cite{Komatsu et al.}) of the energy content of the Universe is made up of non-baryonic particles, which interact predominantly through gravity and have moved with non-relativistic velocities since the earliest epochs of structure formation. While this scenario has been very successful in explaining the formation of large-scale structures in the Universe (galaxies, galaxy groups and galaxy clusters), its predictions on subgalactic scales have not yet been observationally confirmed in any convincing way. On the contrary, there are at least two features of current CDM simulations that appear to be in conflict with empirical data: the existence of high-density cusps in the centres of dark matter halos, and a rich spectrum of substructures within each halo. There is, however, no consensus on how serious these problems really are for the CDM paradigm.

Massive CDM halos are assembled hierarchically from smaller halos. As these subunits fall into the potential well of larger halos, they suffer tidal stripping of material which ends up in the smooth dark matter component of the halo that swallowed them. Since this is a process that may take several billion years to complete, many of these smaller halos temporarily survive in the form of subhalos (also known as halo substructures or subclumps) within the larger halo. According to current simulations, around 10\% of the virial mass of a Milky Way-sized CDM halo should be in the form of subhalos at the current epoch \cite{Springel et al.,Zemp et al.}. Naively, one might expect dwarf galaxies to form in these low-mass halos prior to merging, which would result in large numbers of satellite galaxies within the CDM halo of each large galaxy. A long-standing problem with this picture is that the number of subhalos predicted by simulations greatly exceeds the number of dwarf galaxies seen in the the vicinity of large galaxies like the Milky Way and Andromeda \citep{Klypin et al.,Moore et al. a}. This has become known as the ``missing satellite problem''. A similar lack of dwarf galaxies compared to the number of dark halos predicted is also evident within group-sized dark matter halos \cite{Tully et al.}. While most studies have focused on the discrepancy between the number of subhalos and observed satellite galaxies, the discrepancy persists in the field population as well. In a sense, the missing satellites are just one aspect of a more general problem -- the mismatch between the low-mass end of the dark matter mass function and the luminosity function of dwarf galaxies \cite{Verde et al.}.

A number of potential solutions to the missing satellites problem have been suggested in the literature. These can be sorted into three different categories, depending on how they propose to answer the question ``Do subhalos exist in the numbers predicted by CDM simulations?'': 
\begin{itemize}
\item {\bf No.} In the first category, we find modifications of the properties of dark matter that act to reduce the numbers of low-mass halos and subhalos, including warm dark matter \cite{Bode et al.}, self-interacting dark matter \cite{Spergel & Steinhardt}, fuzzy dark matter \cite{Hu et al.} and dark matter in the form of superWIMPs \cite{Cembranos et al.}, but also models of inflation that produce the required cut-off in the primordial density fluctuation spectrum \cite{Kamionkowski & Liddle}.
\item {\bf Yes.} Here, we find processes for inhibiting star formation in low-mass halos \cite{Bullock et al.,Somerville,Benson et al.,Kravtsov et al.,Moore et al. b} and observational biases that would put the resulting ``dark galaxies'' \cite{Trentham et al.,Verde et al.} outside the reach of current surveys \cite{Simon & Geha,Walsh et al.,Tollerud et al.,Koposov et al. b,Maccio et al. b}. While these mechanisms may be able to solve the missing satellites problem as it was originally defined, solutions of this type imply that a vast population of low-mass CDM subhalos (hosting very faint stellar populations or none at all) should still be awaiting discovery. 
\item {\bf Yes, but not in our neighbourhood.} The final possibility is that the large halo-to-halo scatter in subhalo mass fraction may have left the Milky Way and Andromeda sitting inside CDM halos with unusually few subhalos compared to the cosmic average \cite{Ishiyama et al. a,Ishiyama et al. b}. This would imply that large numbers of CDM subhalos (either bright or dark) should be awaiting discovery in the vicinity of more distant galaxies.
\end{itemize}

Gravitational lensing may play an important role in the quest to determine which of these different solutions is the correct one. If subhalos do exist, lensing can in principle be used to detect even those that are too faint to be observed through other means. If subhalos do not exist, the absence of subhalo-induced lensing effects should be able to tell us so. The goal of this review is to explain how this can be achieved and to point out some potential pitfalls along the way. The material will be described at a level comprehensible even for beginning PhD students, with focus on the big picture rather than the computational details of lensing.

In section 2, we review the properties of the CDM subhalo population, as inferred by current N-body simulations. Section 3 outlines the four expected effects of lensing by CDM subhalos -- flux ratio anomalies, astrometric effects, small-scale image distortions and time delay effects -- which are then covered in more detail in sections 4--7. A number of open questions and future prospects are discussed in section 8.

\section*{2. The properties of cold dark matter subhalos}
N-body simulations indicate that the subhalos within a galaxy-sized CDM halo follow a mass function of the type:
\begin{equation}
\frac{\mathrm{d}N}{\mathrm{d}M_\mathrm{sub}}\propto M_\mathrm{sub}^{-\alpha},
\label{subhalo mass function}
\end{equation}
with $\alpha\approx 1.9$ \cite{Gao et al.,Springel et al.}, albeit with non-negligible halo-to-halo scatter at the high mass end ($M_\mathrm{sub}\gtrsim 5\times 10^8 M_\odot$) \cite{Ishiyama et al. b,Springel et al.}. Current simulations of entire galaxy-sized dark matter halos can resolve subhalos with masses down to $M_\mathrm{sub}\sim 10^5 \ M_\odot$, but the mass function may extend all the way down to the cut-off in the density fluctuation power spectrum, which is set by the detailed properties of the CDM particles. For many types of WIMPs (e.g. neutralinos) this cut-off lies at $\sim 10^{-6} M_\odot$ \cite{Green et al.,Loeb & Zaldarriaga,Diemand et al. a,Profumo et al.,Berezinsky et al.,Bringmann}, but other CDM candidates may alter this truncation mass considerably. As an example, axions may allow the existence of halos with masses as low as $10^{-12}\ M_\odot$ \cite{Hogan & Rees}), whereas very few halos with masses below $10^4$--$10^7\ M_\odot$ are expected in the case of MeV mass dark matter \cite{Hooper et al.}. The overall mass contained in resolved subhalos (i.e $M_\mathrm{sub}\gtrsim 10^5 \ M_\odot$) within a galaxy-sized CDM halo amounts to a subhalo mass fraction around $f_\mathrm{sub}\approx 0.1$, and extrapolating the mass function given by eq. (\ref{subhalo mass function}) towards lower masses does not boost this by much \cite{Springel et al.}.

Since subhalos are more easily disrupted in the central regions of their parent halo, the subhalo population tends to be less centrally concentrated than the smooth CDM component. The spatial distribution of subhalos within $r_{200}$, the radius at which the density of the halo drops below 200 times the critical density of the Universe, can be described by \cite{Madau et al.}:
\begin{equation}
N(<x)=N(<r_{200})\frac{12x^3}{1+11x^2}, 
\end{equation}
where $x=r/r_{200}$ and $N$ denotes the number of subhalos within a specific radius. It should be noted that this result is based on CDM-only simulations, and that the presence of baryons within the subhalos may make them more resistant to tidal disruption, thereby boosting their number densities in the inner regions of their parent halos \cite{Weinberg et al.}. Some simulations of cluster-mass halos have indicated that the spatial distribution of subhalos may be a function of subhalo mass, in the sense that high-mass subhalos would tend to avoid the central regions more than low-mass ones \cite{Ghigna et al.,de Lucia et al.}, but this has not been confirmed by the latest simulations of galaxy-sized halos \cite{Ludlow et al.,Springel et al.}. 

While the term subhalo is typically used to denote clumps located within the virial radius (or, alternatively, $r_{200}$) of a large CDM halo, there is also a large number of low-mass clumps located just outside this limit \cite{Diemand et al. b, Ludlow et al.}. Some of these have previously been bona fide subhalos, and others are bound to venture inside the virial radius in the near future. Such objects can through projection appear close to lines of sight passing through the centres of large galaxies, and may therefore be important in certain lensing situations.

The internal structure of subhalos is still a matter of much debate. As low-mass halos are accreted by more massive ones and become subhalos, substantial mass loss occurs, preferentially from their outer regions. The shape of the outer part of the subhalo density profile may therefore be seriously affected by stripping, whereas the inner profile is left more or less intact. In many lensing studies, CDM subhalos are considered to be singular isothermal spheres (SIS) or even point masses. This is mainly for simplicity -- the lensing properties of such objects are well-known, but neither observations, theory nor simulations favour models of this types for the subhalos predicted by CDM (see \cite{Zackrisson et al.} for references).

An SIS has a density profile given by:
\begin{equation}
\rho_\mathrm{SIS}=\frac{\sigma_\mathrm{v}^2}{2\pi G r^2},
\label{SIS}
\end{equation}

where $\sigma_\mathrm{v}$ is the line-of-sight velocity dispersion. This model has proved to be successful for the massive galaxies responsible for strong lensing (see section 3.1) on arcsecond scales \cite{Rusin et al.,Koopmans et al. a}. The SIS density profile has a steep inner slope ($\rho \propto r^\beta$ with $\beta=-2$), which in the case of massive galaxies is believed to be due to the luminous baryons residing in their inner regions. This baryonic component contributes substantially to the overall mass density in the centre, and its formation over cosmological time scales may also have caused the CDM halo itself to contract, thereby steepening the inner slope of its density profile \cite{Gnedin et al.,Maccio et al. a,Gustafsson et al.,Kampakoglou}. 

Low-mass CDM halos which never formed many stars are unlikely to have density profiles this steep. Instead, they should resemble the halo density profiles derived from CDM-only simulations. The NFW density profile \cite{NFW}, with an inner slope of $\beta=-1$, has for a number of years served as the standard density profile for CDM halos, and is given by: 
\begin{equation}
\rho_\mathrm{NFW}(r)=\frac{\rho_\mathrm{i}}{(r/r_\mathrm{S})(1+r/r_\mathrm{S})^{2}}, 
\end{equation}
where $r_\mathrm{S}$ is the characteristic scale radius of the halo and $\rho_\mathrm{i}$ is related to the density of the Universe at the time of collapse. Modifications of this formula are required once halos become subhalos and are tidally stripped. Attempts to quantify the effects of this mass loss on their density profile have been made by Hayashi et al. \cite{Hayashi et al.} and Kazantzidis et al. \cite{Kazantzidis et al.}

Controversy remains, however, over whether the NFW profile gives the best representation of CDM halos (and, consequently, of subhalos prior to stripping). Based on recent high-resolution simulations, some have argued for a slightly steeper inner slope ($\beta \approx -1.2$; \cite{Diemand et al. c}) with significant halo-to-halo variations, whereas other have favoured a far shallower inner slope \cite{Navarro et al. a,Stadel et al.,Springel et al.,Navarro et al. b}. Inner density profiles as steep as that of the SIS model (\ref{SIS}) are, however, unanimously ruled out. While the internal structure of subhalos may be relatively unimportant in certain lensing situations, it can be crucial in others \cite{Moustakas et al. 09}. For lensing tests that are sensitive to the exact slope of the inner density profile of subhalos \cite{Zackrisson et al.}, the subhalo-to-subhalo scatter in this quantity may also be a very important.  
\begin{figure*}[t]
\resizebox{8.5cm}{!}{\includegraphics{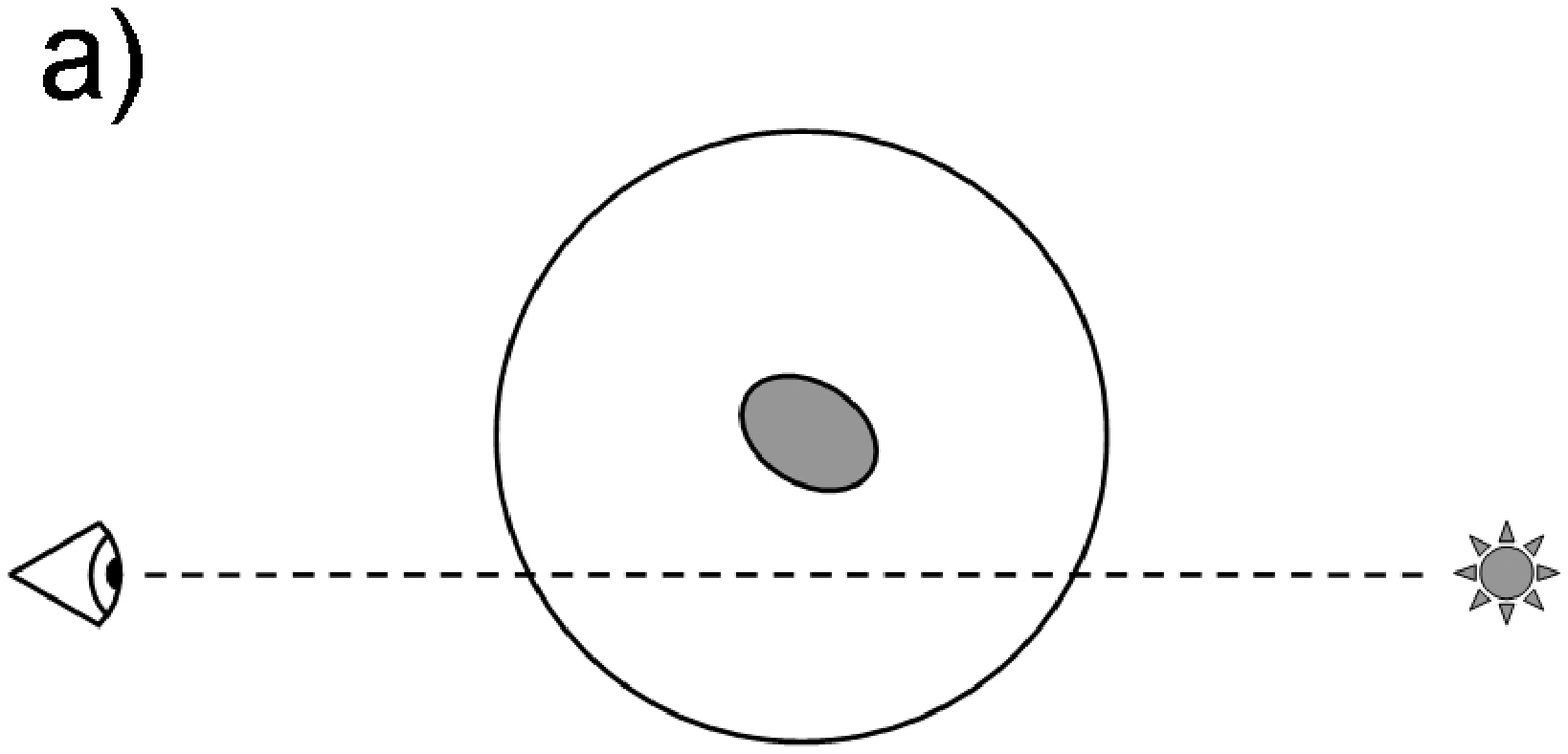}}
\resizebox{8.5cm}{!}{\includegraphics{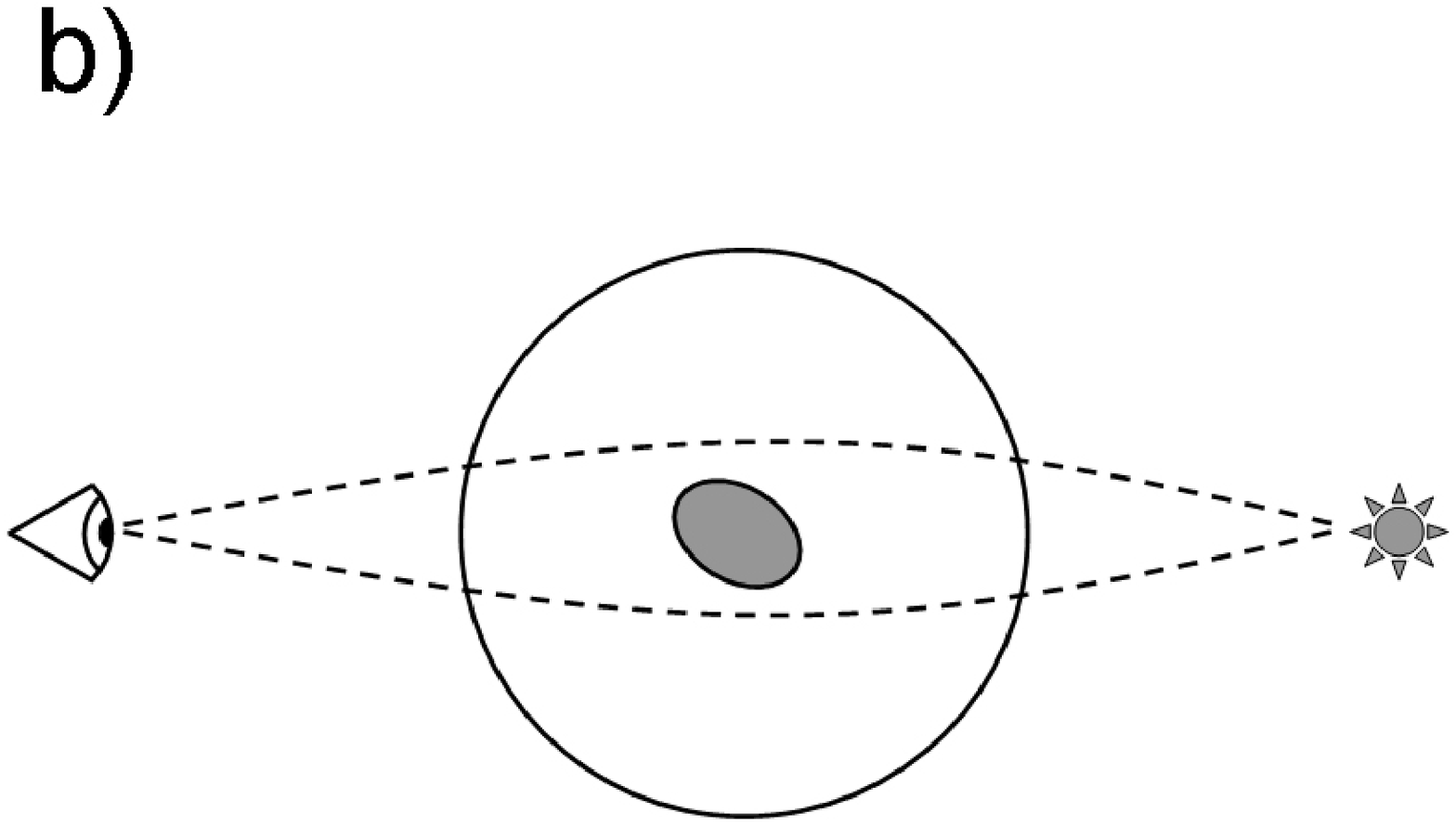}}
\caption{Weak and strong lensing. {\bf a)} Weak lensing occurs when the lens (here illustrated by a gray elliptical galaxy, surrounded by a dark matter halo which extends to the outer circle) lies relatively far from the line of sight between the observer (eye) and the background light source (star). In this case, where the sightline misses the central galaxy, only a single image is produced, subject to mild magnification and distortion. The signatures of this are only detectable in a statistical sense, by studying the weak lensing effects on large numbers of background light sources. {\bf b)} Strong lensing can occur when the dense central region of the lens galaxy is well-aligned with the line of sight. The light from the background light source may then reach the observer along different paths, corresponding to separate images in the sky. This case is also associated with high magnifications and strong image distortions. The angular deflection in this figure, as in all subsequent ones, has been greatly exaggerated for clarity.}
\label{strong vs weak}
\end{figure*}
\begin{figure*}[t]
\resizebox{8.5cm}{!}{\includegraphics{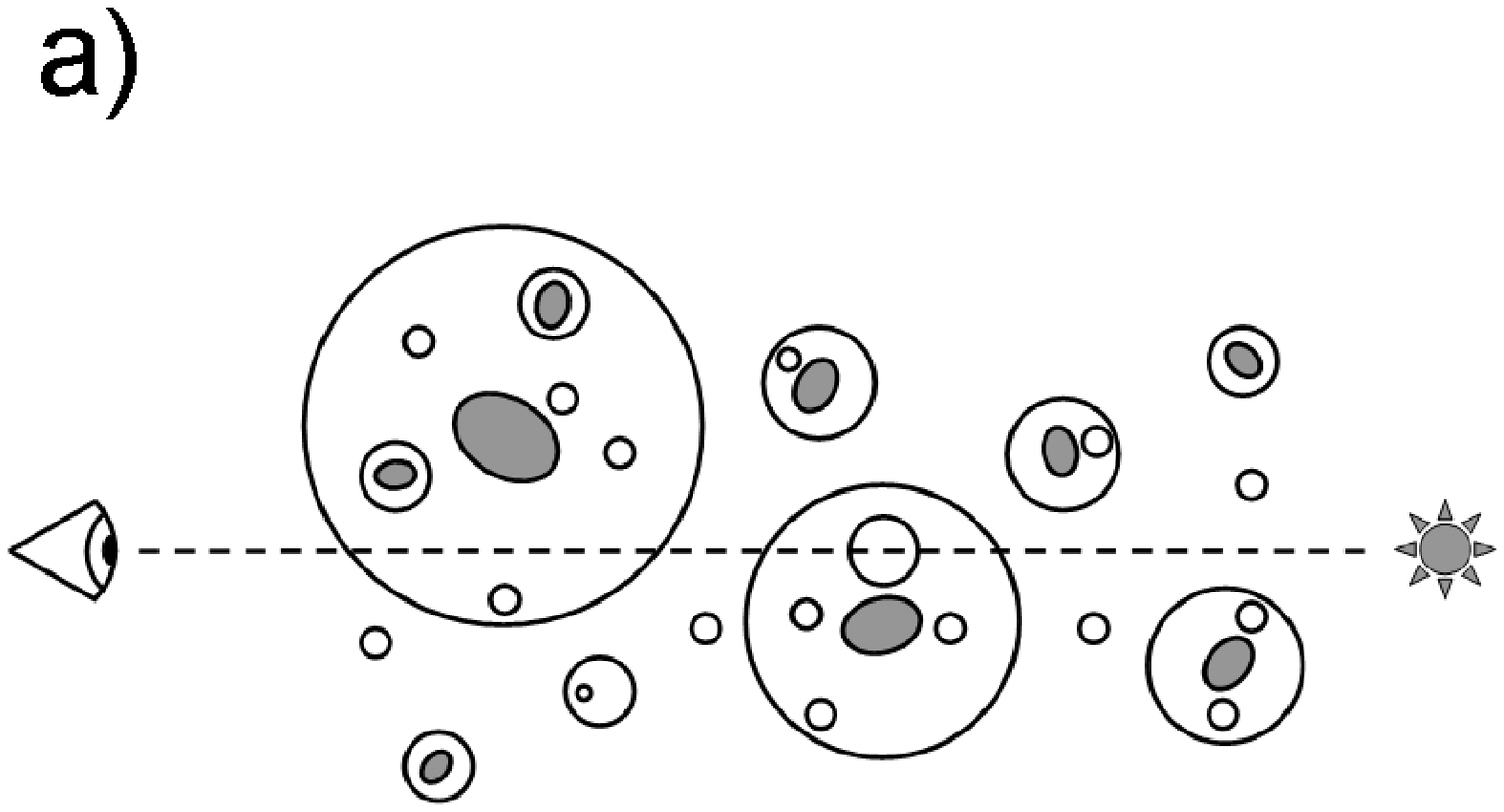}}
\resizebox{8.5cm}{!}{\includegraphics{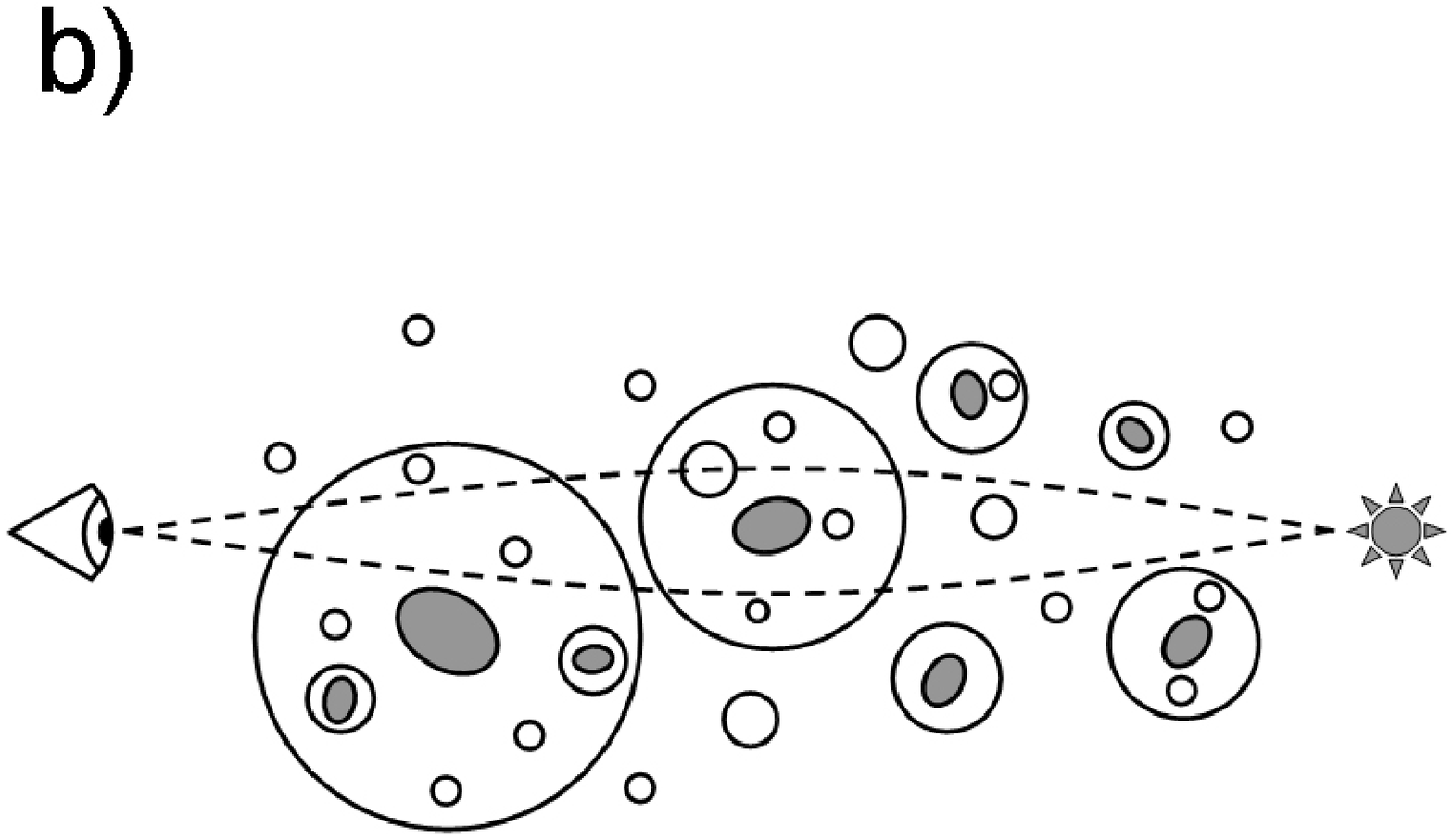}}
\caption{Macrolensed and singly-imaged sources. {\bf a)} The sightline towards a distant light source passes through many dark matter halos with subhalos, but too far from the dense galaxies in the centres of halos for macrolensing to occur. A subhalo in one of these halos happens to intersect the line of sight, thereby potentially producing millilensing effects in a singly-imaged light source.
{\bf b)} One of the halos happens to lie almost exactly on the line of sight, thereby splitting the background light source into separate macroimages. Furthermore, one of the subhalos in the main lens lies on the sightline towards one of the macroimages, thereby producing millilensing effects in this macroimage.}
\label{single vs multiple}
\end{figure*}

\section*{3. Gravitational lensing by cold dark matter subhalos}
The majority of methods aiming to probe CDM subhalos through gravitational lensing focus on detecting subhalos outside the Local Group 
(typically at redshifts $z\approx 0.5$--1.0). Whereas the subhalos sitting within the dark matter halo of the Milky Way will also give rise to lensing, these effects would be very difficult to detect and separate from other phenomena. It has been suggested that CDM subhalos around the Milky Way may be detectable through the gravitational time delay they impose on millisecond pulsars \cite{Siegel a,Siegel b}, but the time scales and probabilities for such events indicate that this is only viable for the very low-mass end of the subhalo mass function ($M<1\ M_\odot$). In what follows, we will therefore focus on the lensing situations relevant for subhalos at cosmological distances.

\subsection*{3.1. Strong and weak lensing}
Gravitational lensing effects can be divided into two regimes, strong and weak lensing, depending on the level of image distortions produced. Strong lensing can occur when the surface mass density along a given sightline exceeds a certain (redshift-dependent) critical value, and is associated with high magnifications, multiple images, arcs and rings in the lens plane. In practice, strong lensing tends to take place when the line of sight from the observer to source lies very close to the centre of a galaxy or a galaxy cluster, since it is only there that sufficiently high surface mass densities are attained. Weak lensing occurs in regions of subcritical surface mass density -- in practice when the lens is located further away from the line of sight -- and gives rise to small magnifications and mild image distortions. The two situations are schematically illustrated in Fig.~\ref{strong vs weak}.

Generally speaking, weak lensing is extremely common in the cosmos (at some level, every single light source is affected) but inconspicuous, and can only be detected statistically by studying a large number of lensed light sources. Strong lensing effects, on the other hand, are rare but dramatic, and can readily be spotted in individual sources. All published strategies for detecting CDM subhalos (in the mass range relevant for the missing satellite problem) through lensing belong to the strong lensing category.

\subsection*{3.2. Macrolensing and millilensing}
Strong lensing can be further divided into subcategories, depending on the typical angular separation of the multiple images produced: macrolensing ($\gtrsim 0.1$ arcseconds), millilensing ($\sim 10^{-3}$ arcseconds), microlensing, ($\sim 10^{-6}$ arcseconds), nanolensing ($\sim 10^{-9}$ arcseconds) and so on. When large galaxies ($M\sim 10^{12}\ M_\odot$) are responsible for the lensing, the image separation typically falls in the macrolensing range, whereas individual solar-mass stars give image separations in the microlensing regime. Since all objects with resolved multiple images due to gravitational lensing have image separations of $\gtrsim 0.1$ arcseconds, the term strong lensing is often used synonymously with macrolensing. Lensing by objects of dwarf-galaxy masses ($\sim 10^6$--$10^{10}\ M_\odot$), like the CDM subhalos relevant for the missing satellite problem, have been estimated to give rise to millilensing, although the exact image separation depends on the internal structure of such objects. In this paper, we will for simplicity use the term millilensing for {\it all} lensing effects associated with dwarf-galaxy mass subhalos, regardless of whether multiple images are produced or not. The term mesolensing is sometimes also used to denote this type of lensing (with angular scales intermediate between microlensing and macrolensing), but since this word also has an alternative meaning in the gravitational lensing literature \cite{Di Stefano}, we will avoid it here.  
\begin{figure*}[t]
\resizebox{8.5cm}{!}{\includegraphics{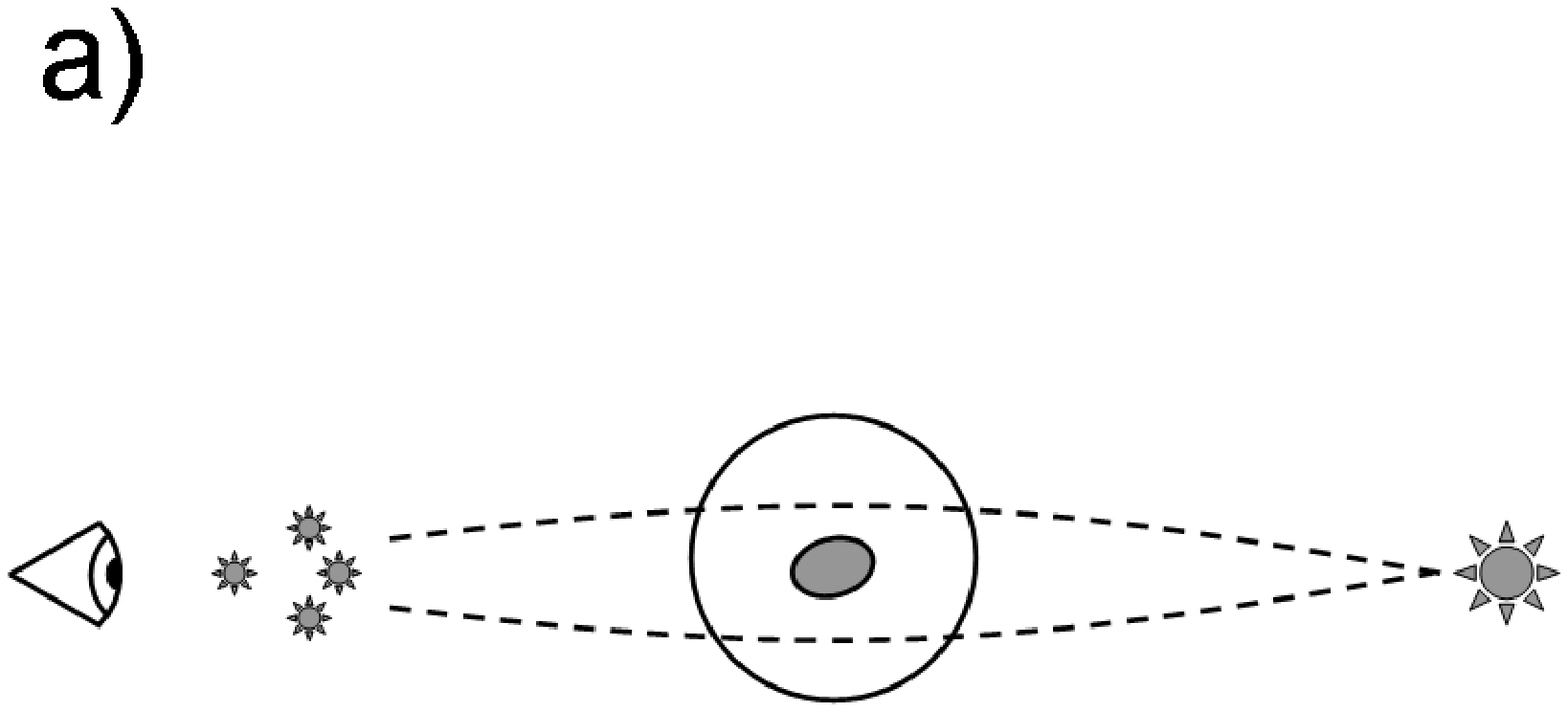}}
\resizebox{8.5cm}{!}{\includegraphics{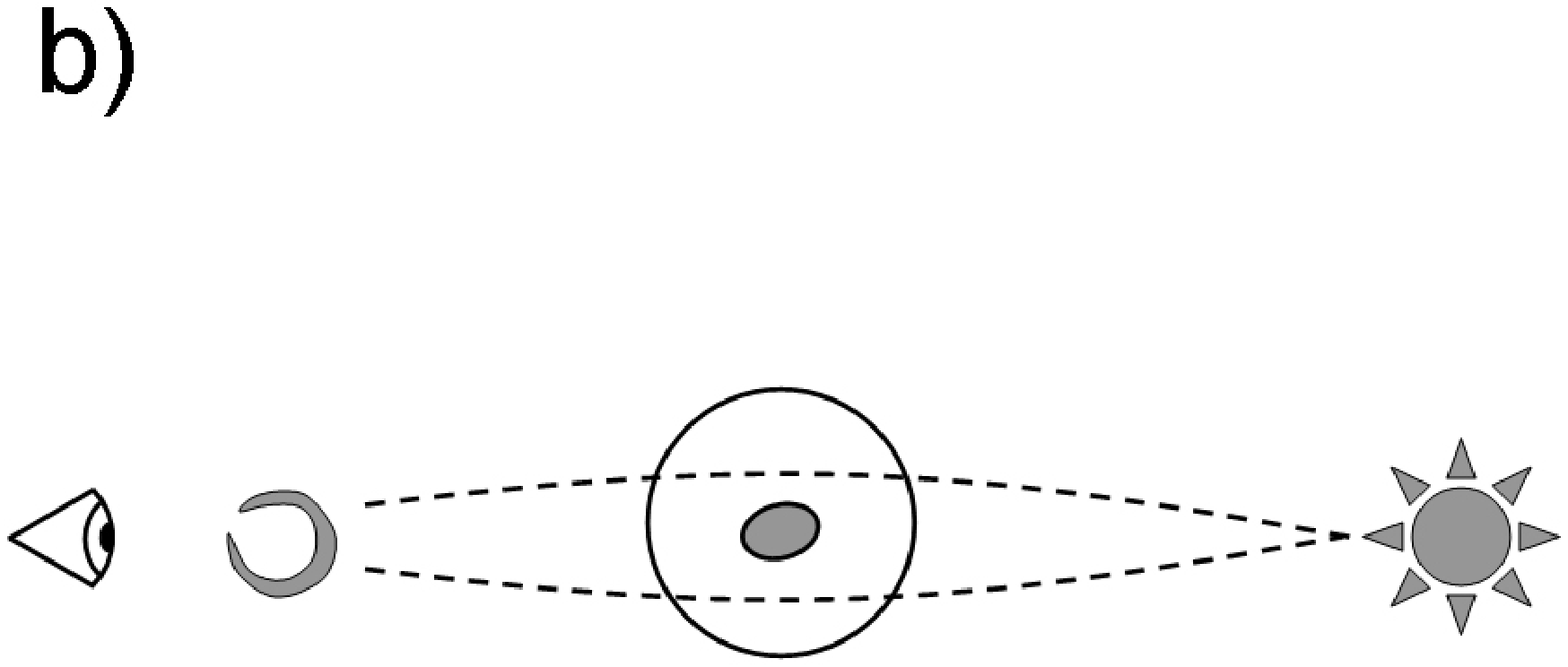}}
\caption{Small and large sources. {\bf a)} A galaxy surrounded by a dark matter halo produces multiple images of a small background light source (e.g. an optical quasar). {\bf b)} For a larger background source (e.g. a galaxy or a radio-loud quasar), the macroimages may be stretched into arcs or even a complete Einstein ring.}
\label{Rings}
\end{figure*}

\subsection{3.3. Suitable light sources}
In principle, any distant light source may be affected by subhalos along the line of sight. The situation is schematically illustrated in Fig.~\ref{single vs multiple}a. A random line of sight towards a light source outside the local volume passes within the virial radius of numerous galaxy-sized CDM halos \cite{Zackrisson & Riehm}, and may hence intersect subhalos anywhere along the line of sight. The probability of hitting a subhalo is, however, rather small in this situation, and sightlines of this type may also pass through low-mass field halos (i.e. the progenitors of subhalos) \cite{Chen et al. b,Metcalf b,Metcalf c,Miranda & Maccio}. It will therefore be difficult to distinguish between the lensing effects produced by these two types of lenses. While the low-mass end of the field halo population may be very interesting in its own right, lensing by such objects is more often considered an unwanted ``background'' when attempting to address the missing satellite problem as it is currently defined. It may also be hard to distinguish complicated intrinsic source structure from extrinsic distortions due to lensing (by either subhalos or low-mass field halos) in this case.

Instead, the main targets for attempts to constrain the CDM subhalo population using lensing have so far been sources that are already known to be macrolensed (see Fig.~\ref{single vs multiple}b), which in practice means observing either multiply-imaged quasars or galaxies lensed into arcs or Einstein rings. By doing so, one preselects a sightline where one knows that there is a massive dark matter halo (and supposedly subhalos) located along the line of sight. Whether one sees several distinct, point-like images, or elongated arcs that approach the form of an Einstein ring, mainly depends on the source size: point-like sources (quasars in the optical, but potentially also supernovae, gamma-ray bursts and their afterglows) give distinct images whereas extended sources (galaxies) give rise to arcs and rings (see Fig.~\ref{Rings}). The strong magnification produced by the macrolens (large foreground galaxy plus dark halo) acts to boost the probability for lensing by the subhalo, and typically augments the observable consequences of such secondary lensing. In the case of multiple images, millilensing by subhalos can moreover be distinguished from intrinsic source features, since any structure intrinsic to the source should be imprinted in all macroimages, whereas millilensing effects will be unique to each image. Transient light sources, like supernovae or gamma-ray bursts can in principle also be used for this endeavour, but no macrolensed sources of this type have so far been detected.

Lensing by subhalos can give rise to a number of observable effects, which we describe in the following sections: flux ratio anomalies, astrometric effects, small-scale structure in macroimages and time-delay effects. In what follows, we will focus on subhalos in the mass range from globular clusters to dwarf galaxies ($\sim 10^5$--$10^{10}\ M_\odot$), since current predictions indicate that subhalos at lower masses may be very difficult to detect through lensing effects.
\begin{figure*}[t]
\resizebox{17.5cm}{!}{\includegraphics{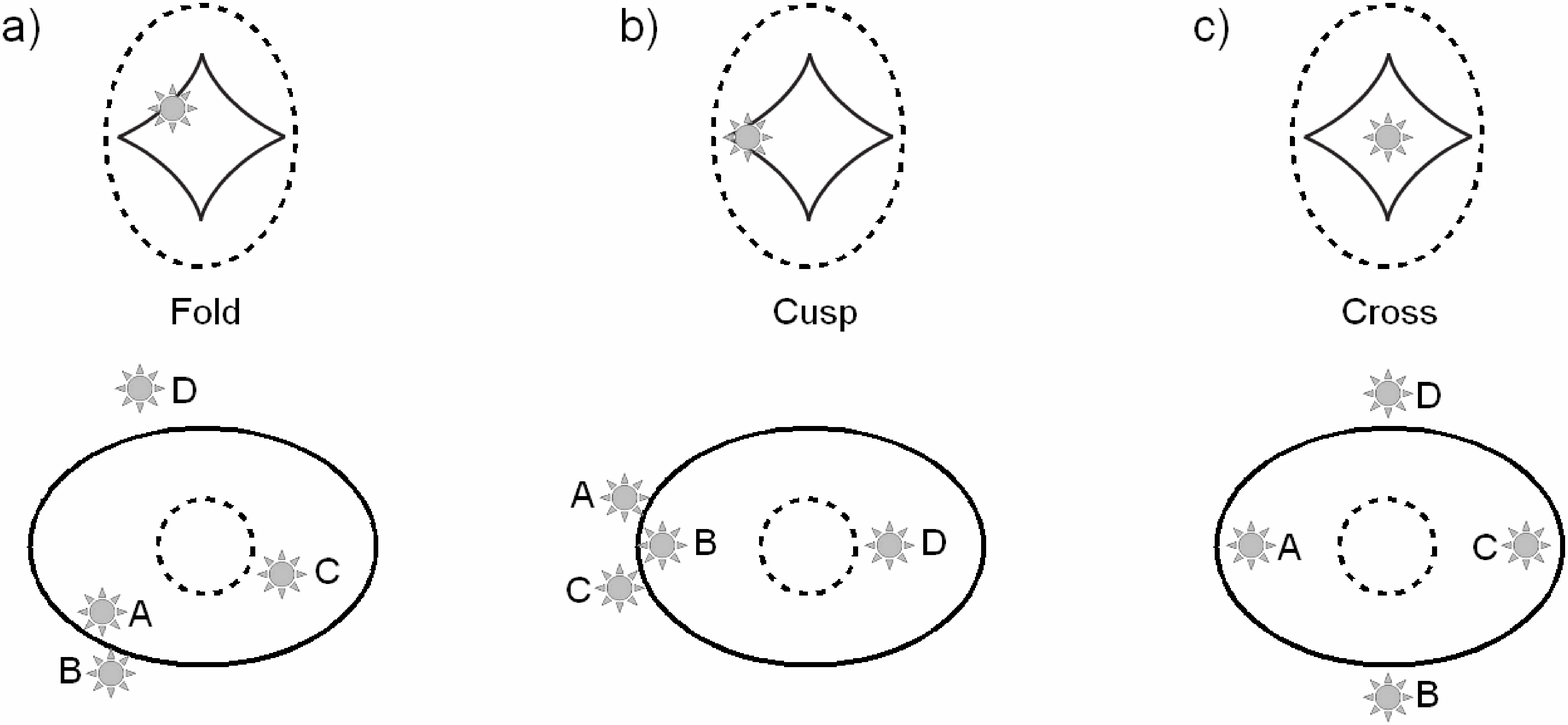}}
\caption{Different configurations of a four-image lens: {\bf a)} Fold, {\bf b)} Cusp and {\bf c)} Cross. The upper row shows the caustics and position of the source (star) in the source plane. The solid line indicates the inner caustic and the dashed line the outer caustic. A source positioned inside the inner caustic produces five images. A source positioned between the inner and outer caustic produces three images, whereas a source positioned outside the outer caustic will not be multiply imaged. In the case of multiple images, one of the images is usually highly de-magnified, so that only four- and two-image lens systems are observed, respectively. The lower row shows the corresponding critical lines and resulting observable images in the lens plane. The inner caustic maps on the outer critical line and vice versa. A close pair (A, B) and a close triplet (A, B, C) are produced in the fold ({\bf{}a}) and cusp ({\bf{}b}) configurations, respectively.}.
\label{Caustics}
\end{figure*}

\section*{4. Flux ratio anomalies}
It was noticed quite early that simple, smooth models of galaxy lenses usually fit the image positions of macrolensed systems well, whereas the magnifications of the macroimages are more difficult to explain \cite{Kochanek}. To see how this works, a bit of simple lens theory is required. 

Specific relations are expected to apply for the magnifications of macroimages close to each other and a critical line. Formally, critical lines are the curves in the lens plane where the magnification tends to infinity. If critical curves are mapped into the source plane, a set of caustic curves is obtained. These separate regions in the source plane that give rise to different numbers of images (see Fig.~\ref{Caustics}). The smooth portions of a caustic curve are called folds, while the points where two folds meet are referred to as cusps. For a background source which is close to either a fold (Fig.~\ref{Caustics}a) or a cusp (Fig.~\ref{Caustics}b) in the caustic of a smooth lens, two respectively three close images will be produced near the critical line in the lens plane. If the source is placed in the center of the caustic, the macroimages will form a cross configuration (Fig.~\ref{Caustics}c). 

All macroimages can furthermore be described as having either positive parity (meaning that the image has the same orientation as the source) or negative parity (the image is mirror flipped in one dimension relative to the source). When taking the image parity into account and assigning negative magnifications to negative parity images, the sum of the magnifications of the close images should approach zero \citep{Mao 92, Schneider & Weiss 92, Zakharov 95}. 
The following relations should then apply for the flux ratio $R$ of a fold configuration:
\begin{equation}
R_{\mathrm{fold}} = \frac{\left|\mu_\mathrm{A}\right|-\left|\mu_\mathrm{B}\right|}{\left|\mu_\mathrm{A}\right|+\left|\mu_\mathrm{B}\right|} \rightarrow 0,
\end{equation}
when the separation between the close images (A \& B in Fig.~\ref{Caustics}a) is asymptotically small \citep{Blandford & Narayan}. Here, $\mu$ represents the magnification of a specific image. For the cusp configuration (Fig.~\ref{Caustics}b), the corresponding relation is:
\begin{equation}
R_{\mathrm{cusp}} = \frac{\left|\mu_\mathrm{A}\right|-\left|\mu_\mathrm{B}\right|+\left|\mu_\mathrm{C}\right|}{\left|\mu_\mathrm{A}\right|+\left|\mu_\mathrm{B}\right|+\left|\mu_\mathrm{C}\right|} \rightarrow 0.
\end{equation}

However, most observed lensing systems violate these relations. This has been interpreted as evidence of small-scale structure in the lens on approximately the scale of the image separations between the close images. Magnifications of individual macroimages due to millilensing by subhalos would indeed cause the values for $R_{\mathrm{fold}}$ and $R_{\mathrm{cusp}}$ to differ from zero fairly independently of the form of the rest of the lens \citep{Mao & Schneider 98, Metcalf & Madau 01, Chiba 02, Dalal & Kochanek 02, Metcalf & Zhao 02, Keeton et al. 03, Kochanek & Dalal 04,Bradac et al.}. 

A notable problem with this picture is that both semi-analytical structure formation models and high-resolution $\Lambda$CDM simulations seem to be unable to reproduce the observed flux ratio anomalies, since the surface mass density in substructure is lower than that required \citep{Zentner & Bullock,Amara et al. 06,Maccio et al. a, Maccio & Miranda 06, Xu et al. 09}. 

\subsection*{4.1. Complications: Propagation effects and microlensing}
Several alternative reasons for the observed flux anomalies have been discussed, such as propagation effects like absorption, scattering or scintillation in the interstellar medium of the lens galaxy \citep{Mittal et al. 07} and microlensing by stars in the lensing galaxy \citep{Schechter & Wambsganss 02}. Since some sources, like quasars, can exhibit intrinsic flux variations on different timescales, flux ratios may also be difficult to interpret if the time delay between the macroimages (see section 7) is not well known. 

The relevance of propagation effects can be tested by supplementary observations of flux ratios at different wavelengths, since flux losses due to such mechanisms should vary as a function of wavelength. Microlensing by stars can be checked for using long-term monitoring, as this type of lensing is transient and expected to introduce extrinsic variability on the order of months. Millilensing by halo substructure can on the other hand be treated as stationary \citep{Metcalf & Madau 01}. Extended sources (e.g. quasars at mid-infrared and radio wavelengths) should also be far less affected by microlensing than small, point-like sources (quasars in the optical and at X-ray wavelengths). Even though it is often assumed that radio observations of quasars are essentially microlensing-free, some caution should be applied, since substantial short-term microlensing variability is possible in the special case of a relativistic radio jet oriented close to the line of sight. This phenomenon has been detected in at least one multiply-imaged system \citep{Koopmans & de Bruyn 00}. 

Mid-infrared imaging of lenses is attractive since the flux at such wavelengths should be free from differences in extinction among the macroimages, in addition to being free from microlensing by stars due to the extended source size. Such observations can therefore be used to test some of the alternative causes for flux ratio anomalies. Recent studies have used this technique to examine several macrolensed quasars with known flux ratio anomalies in the optical \citep{Chiba et al. 05, Sugai et al. 07, Minezaki et al. 09, MacLeod et al. 09}. The mid-infrared flux ratios of about half of these systems can be fitted with smooth lensing models, which means that only the remaining half of these anomalies are due to millilensing by substructures.

There are also other observational features that argue for substructures as the cause of (at least some) flux ratio anomalies. Negative parity images (so called saddle images -- e.g. the middle image (B) of the close triplet in Fig.~\ref{Caustics}b) are often fainter than predicted by smooth lens models. This is expected from millilensing, as the magnification perturbations induced by substructure lensing have been shown to depend on image parity \citep{Schechter & Wambsganss 02, Bradac et al., Kochanek & Dalal 04, Chen 08}. In contrast, such anomalies cannot be attributed to propagation effects since those should statistically affect all types of images similarly, regardless of their parity. Whether the lensing is due to luminous or dark substructures is, however, a different matter. 

\subsection*{4.2. Luminous substructures}
Luminous substructures have been identified in many of the lens systems with known flux ratio anomalies. Including such substructures in the lens model tends to greatly improve the fit to observations. One example of such a lens system is the radio-loud quadruple quasar B2045+265 \citep{Fassnacht et al. 99} which exhibits one of the most extreme anomalous flux ratios known. Recent deep imaging of this system has revealed the presence of a small satellite galaxy which is believed to cause the flux ratio anomaly \citep{McKean et al. 07}. Nearly half of the lenses detected in the Cosmic Lens All-Sky Survey (CLASS) also display luminous satellite galaxies within a few kpc of the primary lensing galaxy \cite{Xu et al. 09}.  

Recently, there have been studies combining the results from simulations and semi-analytical models of galaxy formation to investigate if luminous dwarf galaxies might be able to explain the frequency of flux ratio anomalies observed \citep{Shin & Evans 08, Bryan et al. 08}. They find that the fraction of luminous satellites in group-sized halos is roughly consistent with the observational data within a factor of two, while the results for galaxy-sized halos seem too low to explain the frequency of luminous satellites within the observed systems. The lensing effect of these luminous dwarf galaxies is also somewhat unclear since most satellites found in the inner regions of larger galaxies are expected to be `orphan' galaxies stripped of their dark matter halos. To investigate this further, higher-resolution simulations involving a realistic treatment of the gas processes are required. Possible explanations for the discrepancy between the expected and observed fraction of luminous satellites include dwarf galaxies elsewhere along the line of sight mistakenly identified as the lens perturber \citep{Metcalf b}. Luminous substructures may moreover be more efficient in producing flux ratio anomalies, since they are likely to be denser than dark substructures due to baryon cooling and condensation \citep{Shin & Evans 08}. 

Projection effects are potentially also important for flux ratio anomalies caused by completely dark substructures as one expects a large amount of line-of-sight structure. Although those structures are less effective than substructures within the lens galaxy in inducing magnification perturbations, the overall effect of line-of-sight clumps may be significant \citep{Chen et al. b, Mao et al. 04, Metcalf b,Miranda & Maccio}.
\begin{figure*}[t]
\resizebox{8.5cm}{!}{\includegraphics{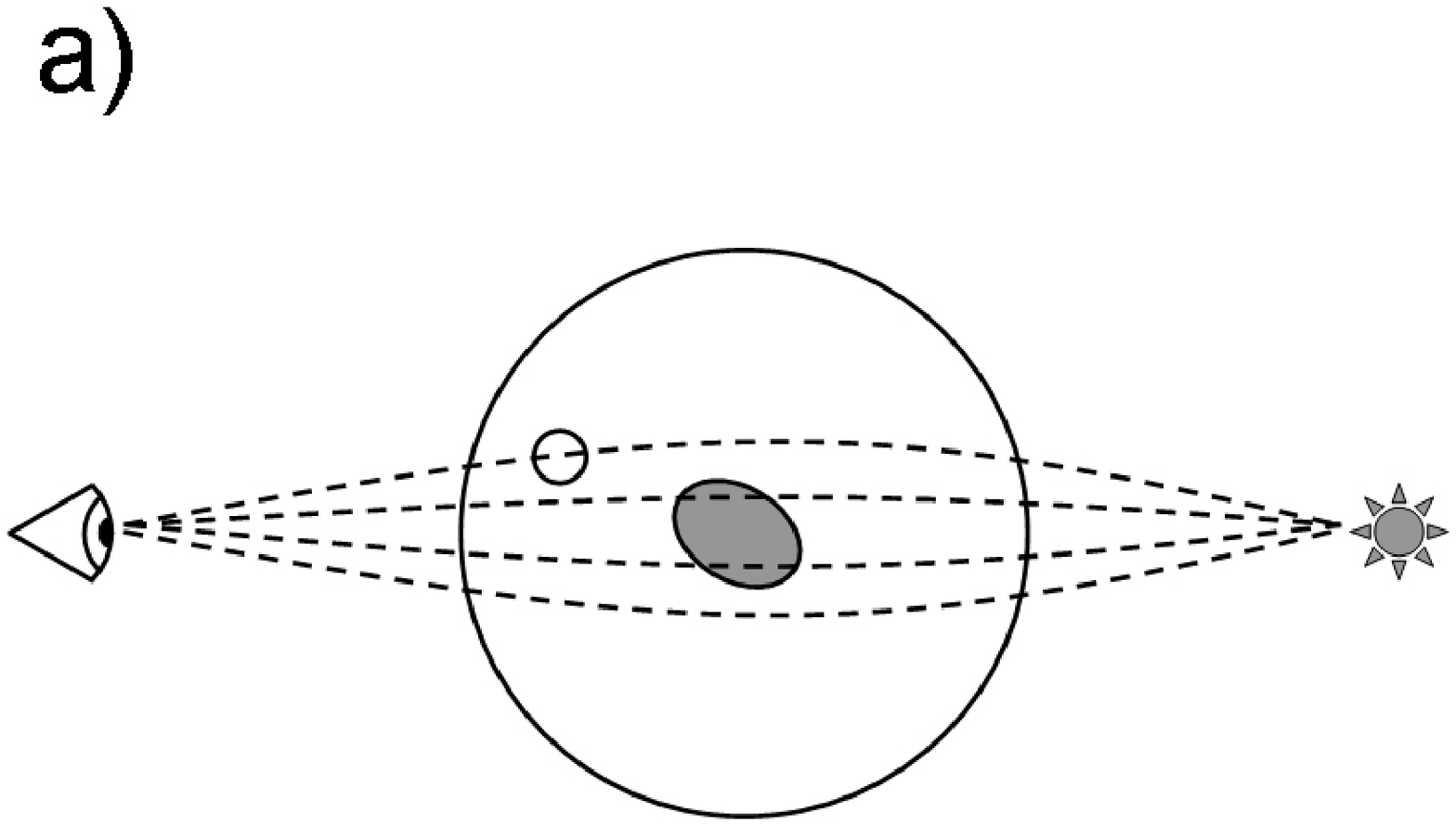}}
\resizebox{8.5cm}{!}{\includegraphics{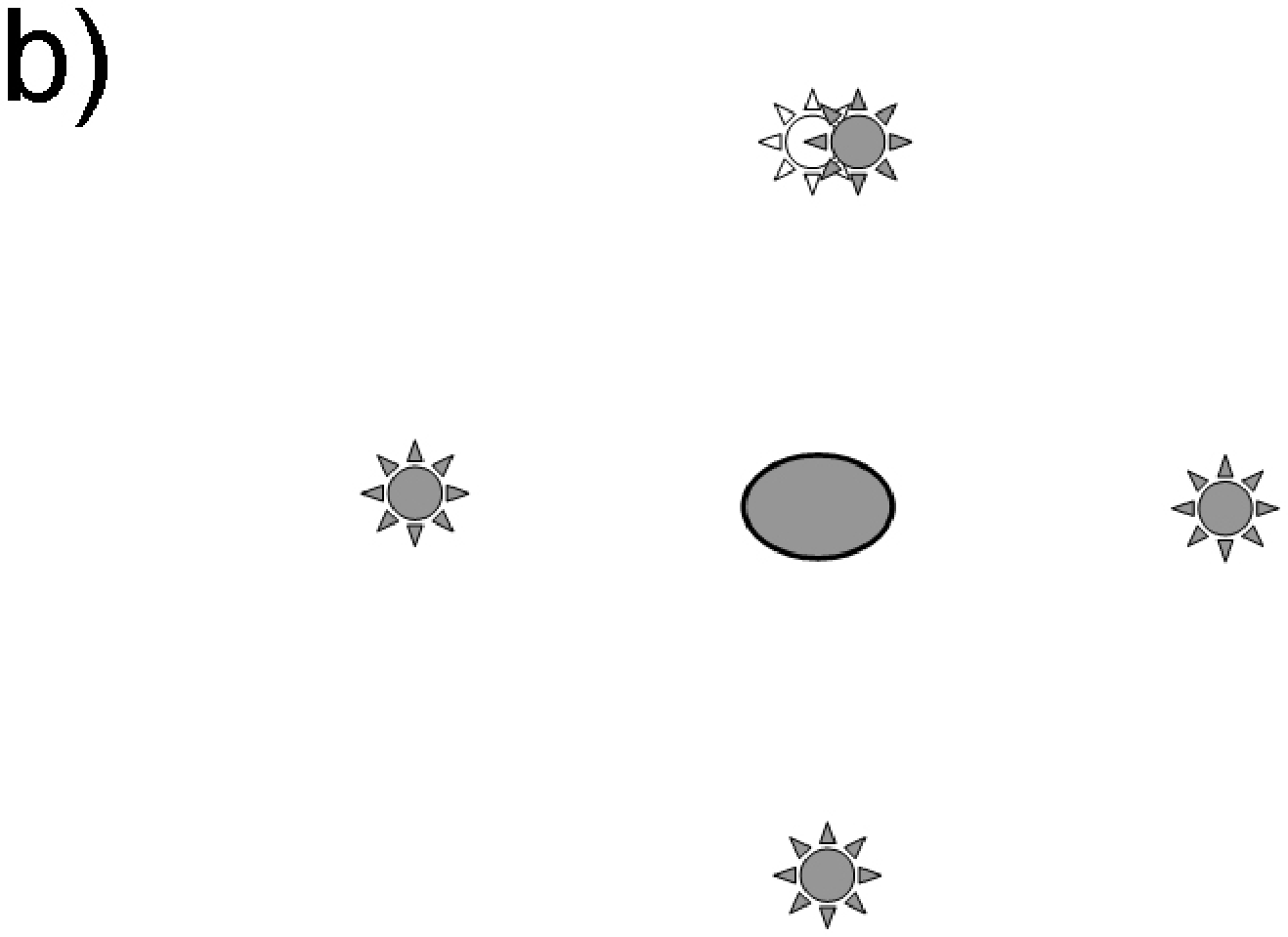}}
\caption{Astrometric perturbations. {\bf a)} One of the multiple sightlines towards a distant light source passes through a dark subhalo. {\bf b)} The images of the macro-lensed source are observed at the positions of the gray source symbols. Modelling of the lens system with a smooth lens potential predicts the position of the upper image at the white source symbol. The subhalo close to the sightline of the image causes a deflection on the order of a few tens of milliarcseconds.}
\label{AstroPer}
\end{figure*}

\section*{5. Astrometric effects}
In macrolensed systems, the presence of halo substructure may perturb the angular deflection caused by the lens galaxy and thereby the position of macroimages at observable levels, so called astrometric perturbations (see Fig.~\ref{AstroPer}). 

This method for detecting subhalos has the advantage of being relatively unaffected by propagation effects (absorption, scattering or scintillation by the interstellar medium) that may contaminate flux ratio measurements. Since the astrometric perturbation is a steeper function of subhalo mass than flux ratio perturbations, it is mostly sensitive to intermediate and high mass substructures and therefore probes a distinct part of the subhalo mass function \citep{Chen et al. 07, Moustakas et al. 09}. Stellar microlensing may complicate the interpretation by producing additional shifts of the positions of macroimages \citep{Williams & Saha}, but such shifts would be transient and predominantly affect point-like sources. 

However, the overall size and probability of subhalo-induced astrometric perturbations are expected to be rather small. Metcalf \& Madau \citep{Metcalf & Madau 01} used lensing simulations of random realizations of substructure in regions near images and found that it would take substructures with masses $\gtrsim 10^8$ M$_{\odot}$ that are very closely aligned with the images to change the image positions by a few tens of milliarcseconds. Such an alignment would be rare in the CDM model. Therefore, they suggest to employ lensed jets of quasars observed at radio wavelengths, as such sources would cover more area on the lens plane. This would increase the probability of having a large subhalo nearby, but still allow for pronounced distortions due to the thinness of the jet. Metcalf \citep{Metcalf 02} investigated this technique further and used it to show that the lens system B1152+199 is likely to contain a substructure of mass $\sim 10^5 - 10^7 M_{\odot}$.

Further observational evidence for astrometric perturbations from small scale structure was found in the detailed image structures of B2016+112 \citep{Koopmans et al. 02, More et al. 09} and B0123+437 \citep{Biggs et al. 04}. In the latter system, a substructure of at least $\sim 10^6 M_{\odot}$ would be needed in order to reproduce the observed image positions.

The CDM scenario predicts that there are far more low-mass subhalos than high-mass ones (see equation \ref{subhalo mass function}) and their summed effect could in principle add up to a substantial perturbation. Conversely, since perturbers positioned on opposite sides around the macrolens generate equal but opposite perturbations, the net effect of a large number of substructures may cancel out, ensuring that rare, massive substructures dominate the position perturbation of the images. Chen et al. \citep{Chen et al. 07} have investigated this by modelling the effects of a wide range of subhalo masses and found that all residual distributions have very large peak perturbations ($\gtrsim$ 10 milliarcseconds). Since the simulation models predict extremely few or no substructures in the inner region of the lens, the perturbers must be located further away. Therefore, it was also inferred that position perturbations of different images in any lens configuration may be strongly correlated. Although these results suggested that rare, massive clumps may cause larger perturbations than the more abundant smaller clumps, the astrometric perturbations of the images were considerable even in models where no such massive substructures were present. On the other hand, these perturbations are at least partly degenerate with model parameters of the host halo.

Since astrometric perturbations are expected to manifest themselves at (sub-)milliarcsecond levels, high spatial resolution observations are required which so far are mainly achieved by Very Long Baseline Interferometry (VLBI) observations of radio-loud quasars.
 
However, recent studies have shown that perturbation effects of substructure should also be detectable on larger scales ($\sim$ 0.1 arcseconds) and at shorter wavelengths in extended Einstein rings and arcs produced by galaxy-galaxy lensing. Peirani et al. \citep{Peirani et al. 08} used the perturbative method and lens distributions from toy models as well as cosmological simulations to predict the possible signatures of substructures. They show that  when a substructure is positioned near the critical line, not only astrometric but also morphological effects, i.e. breaking of the image, will occur which are approximately 10 times larger and should be easier to detect.  

Other studies have suggested to use non-parametric source and lens potential reconstructions to probe small perturbations in the lens potential of highly magnified Einstein rings and arcs (e.g. \citep{Blandford et al. 01, Koopmans 05}). Vegetti \& Koopmans \citep{Vegetti & Koopmans 09a} have used an adaptive-grid method and shown that for substructures located on or close to the Einstein ring, perturbations with masses $\gtrsim 10^7 M_{\odot}$ respectively $10^9 M_{\odot}$ can be reconstructed. This technique may then be used to constrain the substructure mass fraction and their mass-function slope, once a larger sample of high-resolution lenses becomes available \citep{Vegetti & Koopmans 09b}.

With the upcoming generation of telescopes -- e.g. the Large Synoptic Survey Telescope (LSST), the Joint Dark Energy Mission (JDEM), the James Webb Space Telescope (JWST) and the Atacama Large Millimeter Array (ALMA) -- a significant increase in the number a macrolensed lenses is expected which will allow the use of astrometric perturbations as a probe of the subhalo population.
\begin{figure*}[t]
\resizebox{8.5cm}{!}{\includegraphics{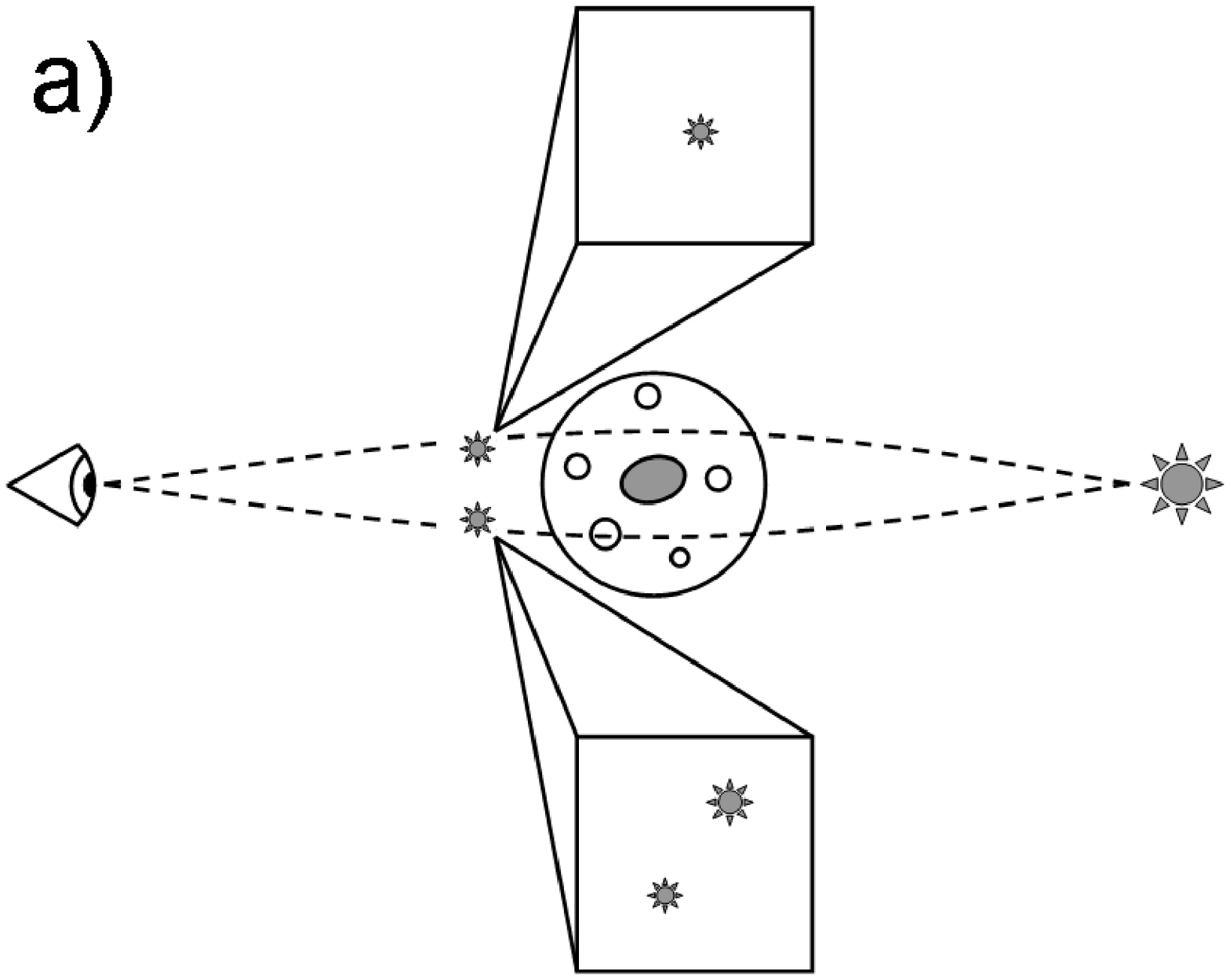}}
\resizebox{8.5cm}{!}{\includegraphics{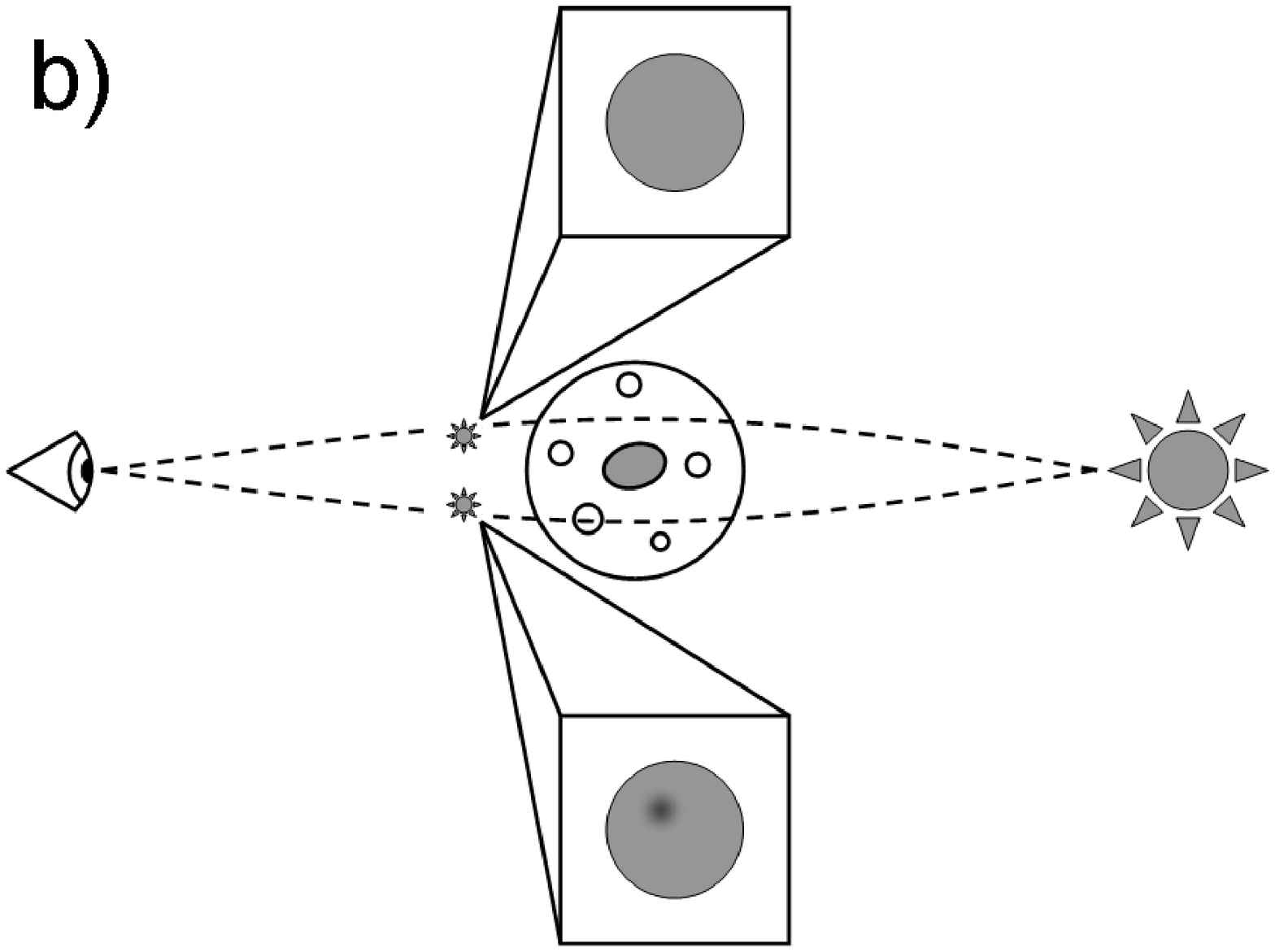}}
\caption{A foreground galaxy with a dark matter halo produces multiple macroimages of a background light source. A subhalo located in the dark halo intercepts one of these macroimages, which may give rise to {\bf a)} additional image small-scale splitting of the affected macroimage if the source is sufficiently small, or {\bf b)} a mild distortion in the affected macroimage, if the source is large.}
\label{Small-scale}
\end{figure*}

\section*{6. Small-scale structure in macroimages}
When dark objects in the dwarf-galaxy mass range intersect the line of sight towards distant quasars, image-splitting or distortion on characteristic scales of milliarcseconds may occur \cite{Kassiola et al.,Wambsganss & Paczynski}. As already mentioned, quasars can currently only be probed on such small scales using VLBI techniques at radio wavelengths, but future telescopes and instruments may allow similar angular resolution at both optical and X-ray wavelengths \cite{Zackrisson et al.}. 

Using VLBI, Wilkinson et al. \cite{Wilkinson et al.} reported no detections of millilensing among 300 compact-radio sources and was able to impose an upper limit of $\Omega<0.01$ on the cosmological density of point-mass objects (i.e., very compact objects, like black holes) in the $10^6$--$10^8\ M_\odot$ range. However, this does not convert into any strong limits on the subhalo population, since CDM halos and subhalos are not nearly as dense as black holes. Correcting for this would decrease the expected image separations for a millilens of a given mass and yield a probability for lensing that is much lower than assumed in their analysis. The sources used were moreover not macrolensed -- this would have made it difficult to make the distinction between subhalos and low-mass field halos as the main culprits even if any signs of millilensing had been detected (see Fig.~\ref{single vs multiple}a).

The effects that a subhalo can have on the internal structure of one of the macroimages in a multiply-imaged quasar (Fig.~\ref{single vs multiple}b) are schematically illustrated in Fig.~\ref{Small-scale}. For a small, point-like source (e.g. a quasar observed at optical wavelengths), the macroimage may split into several distinct images with small angular separations (Fig.~\ref{Small-scale}a). A larger source (e.g. a quasar at radio wavelengths) may instead exhibit small-scale image distortions (Fig.~\ref{Small-scale}b). Even though quasars may display complicated {\it intrinsic} structure when imaged with high spatial resolution, such effects can at least in principle be separated from the features imprinted by millilensing, since intrinsic structure will be reproduced in all macroimages, whereas millilensing effects are unique to each macroimage \footnotemark\footnotetext{The distinction between these small-scale changes in the morphologies of macroimages, and the astrometric effects discussed in section 5, becomes somewhat arbitrary in some cases since image distortion may both shift the centroid of an image {\it and} alter its overall appearance (e.g. through the introduction of new, small-scale images). The distortion of macrolensed jets is for instance usually referred to as an astrometric effect.}.
 
Yonehara et al. \cite{Yonehara et al.} have argued that a significant fraction of all macrolensed optical quasars  may exhibit secondary image-splitting on milliarcsecond scales due to CDM subhalos. Inoue \& Chiba \cite{Inoue & Chiba a,Inoue & Chiba b} have explored a similar scenario in the case of the extended images expected for macrolensed quasars at longer wavelengths, and concluded that the small-scale macroimage distortions produced by CDM subhalos may be detectable with upcoming radio facilities such as ALMA or the VLBI Space Observatory Programme 2 (VSOP-2).   

The merits of probing CDM subhalos through the small-scale structure of macroimages is that, contrary to the case for flux ratio anomalies, there is little risk of confusion due to  microlensing by stars or propagation effects in the interstellar medium. Globular clusters may be able to produce similar effects \cite{Baryshev & Ezova}, and so may luminous dwarf galaxies (i.e. the subset of CDM subhalos that happen to have experienced substantial star formation), but subhalos are expected to outnumber both of these populations, at least in most mass intervals. Instead, the main problem with this approach seems to be that CDM subhalos may not be sufficiently dense to produce multiple images on scales that can be resolved by current technology. Most studies of these effects have assumed that CDM subhalos can be treated as SIS lenses, resulting in a gross overprediction of the image separations compared to more realistic subhalo models \cite{Zackrisson et al.}. The angular resolution by which macrolensed quasars can be probed is on the other hand likely to increase substantially in the coming years, in principle reaching $\approx 0.04$ milliarcseconds with the VSOP-2 mission (scheduled to launch in 2013).

\section*{7. Time delay effects}
The images of a macrolensed light source (see Fig.~\ref{Rings}a) are subject to different time delays, which become detectable when the source exhibits intrinsic temporal variability over observable time scales. These time delays stem from a combination of differences in the relativistic time delays (clocks running slower in deep gravitational fields, also known as Shapiro time delays) and the differences in photon path lengths (due to geometric deflection) among the macroimages. Since quasars are both non-transient and known to vary significantly in brightness on time scales of hours and upwards, they are very convenient targets for observing campaigns aiming to measure such time delays. At the current time, around 20 macrolensed quasars have measured time delays (with typical delays of $\Delta t\sim0.1$--400 days; see Oguri \cite{Oguri} for a recent compilation). Time delays of this type have often been used to constrain the Hubble constant and the density profile of the macrolens (i.e. the overall gravitational potential of the lens galaxy and its associated dark halo), but can also potentially be used to probe the CDM subhalos of the lens galaxy. 

As shown by Keeton \& Moustakas \cite{Keeton & Moustakas}, the presence of subhalos within the macrolens will perturb the time delays predicted by smooth lens models, and may also violate the predicted arrival-time ordering of the images. Such violations would signal the presence of subhalos in a way that, unlike the case for optical flux ratio anomalies, cannot be mimicked by dust extinction or microlensing by stars. The time delay perturbations due to subhalos are typically on the order of a fraction of a day. By pushing the uncertainties in the observed time delays to this level, strong constraints on CDM subhalo populations may potentially be derived. One case of a time ordering reversal which may possibly be attributed to subhalos has already been identified in the macrolensed quasar RX J1131 - 1231 \cite{Morgan et al.,Keeton & Moustakas}.

\begin{figure}[t]
\resizebox{8.5cm}{!}{\includegraphics{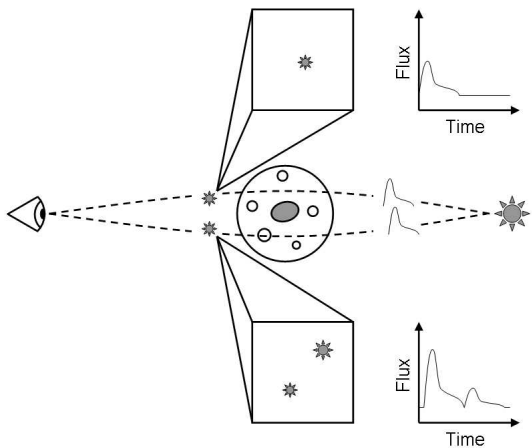}}
\caption{A galaxy with a dark matter halo produces distinct macroimages of a background light source. If this source displays intrinsic variability, observable time delays between the different macroimages may occur. If one of the macroimages experiences further small-scale image splitting due to a subhalo along the line of sight, a light echo may be observable in the affected macroimage. This may serve as a signature of millilensing in cases where the small-scale images blend into one due to insufficient angular resolution of the observations.}
\label{Time delays}
\end{figure}

If the subhalos themselves give rise to small-scale image splittings (as described in section 6), short time lags between the light pulses of the separate small-scale images would be introduced. This imprints echo-like signatures in the overall light curve of astronomical objects with short-term variability (such as gamma-ray bursts and X-ray quasars), even if the small-scale images cannot be spatially resolved. These echos correspond to light signals transported through small-scale images with longer time delays than the leading image, and the flux ratios of the peaks are given by the different magnifications of these images. The light curves of gamma-ray bursts have been used to search for such light echos in the interval $\sim 1$--60 s, resulting in upper limits ($\Omega < 0.1$) on point-mass dark objects in the $10^5$--$10^9 \ M_\odot$ range \cite{Nemiroff et al.} and even a few candidate detections of repeating flares due to millilensing \cite{Ougolnikov}. However, just like in the case of the search for spatial millilensing effects by Wilkinson et al. \cite{Wilkinson et al.}, current investigations of this kind have little bearing on CDM subhalos, since the probability for subhalo millilensing is too low when the target objects are not macrolensed. Yonehara et al. \cite{Yonehara et al.} instead suggested monitoring of macrolensed quasars, predicting that CDM subhalos may produce light echos separated by $\sim 1000$ s, which could potentially be detected in X-rays, where rapid intrinsic flares have been observed. This lensing situation is schematically illustrated in Fig.~\ref{Time delays}.

\section*{8. Open questions and future prospects}
As we have argued, lensing can in principle be used to probe the CDM subhalo population, but has so far not resulted in any strong constraints. Most studies have focused on flux ratio anomalies, but a number of studies now suggest that subhalos by themselves are unable to explain this phenomenon \cite{Mao et al. 04,Amara et al. 06,Miranda & Maccio,Maccio & Miranda 06,Xu et al. 09}. If correct, this would limit the usefulness of this diagnostic, since some other mechanism must also be affecting the flux ratios. Luckily, constraints from other techniques, such as astrometric perturbations, small-scale image distortions and time delay perturbations may be just around the corner.  

Observationally, the future for the study of strong gravitational lensing is looking bright. As of 2009, around 200 macrolensed systems have been detected with galaxies acting as the main lens. Planned observational facilities such as the Square Kilometer Array (SKA) and the LOw Frequency ARray for radio astronomy (LOFAR) at radio wavelengths and JDEM \& LSST in the optical have the power to boost this number by orders of magnitude in the coming decade \cite{Koopmans et al. b}. The spatial resolution by which these systems can be studied is also likely to become significantly better, approaching $\sim 10$ milliarcseconds in the optical and $\sim 0.1$ milliarcseconds at radio wavelenghts \cite{Zackrisson et al.}.  

On the modelling side, there are still a number of issues that need to be properly addressed before strong constraints on the existence and properties of CDM subhalos can be extracted from such data. 

\subsection{8.1. Input needed from subhalo simulations}
The largest N-body simulations of galaxy-sized halos are now able to resolve CDM subhalos with masses down to $\sim 10^5\ M_\odot$, but there are still a number of aspects of the subhalo population that remain poorly quantified and could have a significant impact on its lensing signatures:
\begin{itemize}
\item What is the halo-to-halo scatter in the subhalo mass function and how does this evolve with redshift?
\item What are the density profiles of subhalos? How does this evolve with subhalo mass and subhalo position within the parent halo? How large is the difference from subhalo to subhalo?
\item What is the spatial distribution of subhalos as a function of subhalo mass within the parent halo? What is corresponding distribution outside the virial radius?
\item How do baryons affect the properties of subhalos? Can baryons promote the survival of subhalos within the inner regions of their host halos? 
\end{itemize} 

The lensing effects discussed in sections 4--7 are sensitive to the density profiles and mass function of subhalos, albeit to a varying degree \cite{Moustakas et al. 09}. Attempts to quantify the effects of different density profiles of lensing signature have been made \cite{Maccio & Miranda 06,Chen et al. 07,Zackrisson et al.,Keeton & Moustakas,Shin & Evans 08}, but the models used are still far from realistic, and many of those active in this field still cling to SIS profiles for simplicity. 

\subsection{8.2 The role of other small-scale structure}
CDM subhalos are not the only objects along the line of sight to high-redshift light sources that are capable of producing millilensing effects. Many large galaxies are known to be surrounded by $10^2$--$10^3$ globular clusters with masses in the $10^5$--$10^6\ M_\odot$ range. While typically less numerous than CDM subhalos in the same mass range, they are concentrated within a smaller volume (the stellar halo) and have more centrally concentrated density profiles, thereby potentially making them more efficient lenses. We also expect a fair share of luminous dwarf galaxies within the dark halos of large galaxies. These dwarfs may well represent the subset of CDM subhalos inside which baryons were able to collapse and form stars, but if so, this means that they may have density profiles significantly more centrally concentrated than their dark siblings. While the role of globular clusters and luminous satellite galaxies has been studied in the case of flux ratio anomalies \cite{Chiba 02,Shin & Evans 08}, their effects on many of the other lensing situations discussed in previous sections have not yet been addressed. Low-mass halos along the line of sight may also affect these lensing signatures, and sometimes appreciably so \cite{Metcalf b,Miranda & Maccio}. 

Aside from dwarf galaxies and globulars, there may of course be other surprises hiding in the dark halos of galaxies. Intermediate mass ($M\sim 10^2$--$10^4\ M_\odot$) black holes, formed either in the very early Universe or as the remnants of population III stars may inhabit the halo region \cite{van der Marel,Zhao & Silk,Kawaguchi et al.} and could give rise to millilensing effects \cite{Inoue & Chiba c}. If accretion onto such objects is efficient, the predicted X-ray properties of such black holes already place very strong constraints on their contribution to the dark matter ($\Omega\lesssim 0.005$ \cite{Mapelli et al.}), but the more generalized dynamical \cite{Carr & Sakellariadou,Yoo et al.,Quinn et al.} and lensing \cite{Nemiroff et al.,Metcalf & Silk} constraints on other types of dark objects in the star cluster mass range ($\sim 10^2$--$10^5\ M_\odot$) are otherwise rather weak ($\Omega\lesssim 0.1$). Lensing observations originally aimed to constrain the CDM subhalo population may therefore also lead to the detections of completely new types of halo substructure. As telescopes attain better sensitivity and higher angular resolution in the next decade, we can surely look forward to an exciting new era in the study of dark matter halos.

\section*{Acknowledgements}EZ acknowledges support from the Swedish Research Council. The authors are indebted to Robert Cumming for a careful reading of the manuscript and to two anonymous referees for insightful comments.

\end{document}